\documentstyle[12pt]{article}
\begin{document}
\title
{Higgs boson contributions to neutrino production in $e^-e^+$ collisions
in a left-right
symmetric model\thanks{This work was supported by Polish
Committee for Scientific Researches under Grants Nos. 2252/2/91 and
2P30225206/93}}
\author
{J.Gluza $^{\dag}$ and M.Zra\l ek$^{\ddagger}$ \\
Department of Field Theory and Particle Physics \\
Institute of Physics, University of Silesia\\
Uniwersytecka 4, PL-40-007 Katowice, Poland}
\maketitle
PACS number(s): 13.15.-f, 12.15.Cc,14.60.Gh
\begin{abstract}

In gauge models with bigger number of Higgs particles their
couplings to fermions
are more complicated then in the standard
model (SM). The influence of the Higgs
bosons exchange on the neutrino
production cross section in $e^-e^+$ collision
($e^-e^+ \rightarrow \nu N$)
is investigated. The study has been done in the
framework of the
left-right symmetric model
with the rich Higgs sector containing bidoublet and two triplets.
The
couplings of the Higgs particles even to light leptons can be as
large as the gauge
bosons couplings. The total effect of the Higgs
bosons is small as a result of the big Higgs
mass in the propagators.
\end{abstract}
\newpage
\baselineskip 6mm
\section{Introduction}
\ \ \
Despite the extraordinary phenomenological success of the
SM it is not excluded that new gauge interactions will become visible already
at TeV energies. Many such extended gauge theories have been suggested. One
of the most popular is the model with right-handed currents based on the
symmetry group $SU_L(2) \otimes SU_R(2) \otimes U_{B-L}(1) $ \cite{1intr}.
Left and right-handed weak neutrino states appear in the natural way in this
model. In the most popular version of the model with one bidoublet and two
triplets of the Higgs particles \cite{2intr} neutrino mass matrix
diagonalization
gives three heavy and three light physical Majorana neutrinos. It is possible
that the mass of some heavy neutrinos lie in the energy  of the new symmetry
breaking scale. These GeV-TeV mass neutrinos can be produced in the future
colliders. Light neutrinos with the eV-keV-MeV range masses are generated by
`see-saw' mechanism \cite{3intr}.
In the previous paper \cite{4intr} we investigated the production of two
heavy and light-heavy neutrinos in the $e^+e^-$ future colliders (LEP II,
Next Linear Colliders (NLC),...).
Six Feynman diagrams with two charged gauge bosons $W^{\pm}_{1,2}$ in t and u
channels and two neutral bosons $Z_{1,2}$ in s-channel were taken into account.
We didn't consider any diagram with Higgs particles exchange. In the
models with one Higgs particle like in the SM the Higgs-fermions coupling is
proportional to the fermion mass. In the energy range where heavy neutrinos
are produced in $e^+e^-$ scattering the electron mass and thus Higgs exchange
diagrams are negligible. In the models with greater number of Higgs particles,
their coupling to fermions is more complicated. It is not so obvious then
that the Higgs exchange diagrams in the considered process are also negligible.
On the pure phenomenological ground this mechanism in $e^-e^+ \rightarrow \nu
N$
process has been considered in the literature \cite{5intr}. In this paper we
consider the Higgs exchange mechanism in all details. First, in the next
Chapter, the couplings of physical Higgs particles to the physical leptons are
considered. Next (Chapter III) we discuss the parametrization of the neutrino
mass matrix and the various mixing matrices between weak and mass states.
The problem of CP conservation and non conservation is considered. In Chapter
IV
the results of numerical calculations are presented and finally Chapter V
contains our conclusions.
\section{Higgs boson couplings with fermion pairs.}
\ \ \
 To find the influence of the Higgs bosons on the neutrino production process
$e^-e^+ \rightarrow \nu N$ we need to know the Higgs boson couplings with
fermions. All details of the $SU(2)_L \otimes SU(2)_R \otimes U(1)_{B-L}$
left-right symmetric models which we consider are described in Ref.[2]
and \cite{4intr}.
Before the spontaneous symmetry breaking the Lagrangian has the left-right
symmetry specified by the transformation
\begin{equation}
\Psi_L \leftrightarrow \Psi_R \;\;,\;\; \Delta_L \leftrightarrow \Delta_R\;\;\;
\mbox{\rm and}\;\;\;\; \phi \leftrightarrow \phi^{\dagger}
\end{equation}
where $\Psi_{L,R}$ are the column vectors containing the left-handed and right-
handed leptons, $\phi$ and $\Delta_{L,R}$ are bidoublets and
left (right)-handed Higgs triplets respectively
\begin{equation}
\phi= \left( \matrix{ \phi_1^0 & \phi_1^+ \cr
                      \phi_2^- & \phi_2^0 \cr} \right) \;,\;
\Delta_{L,R}= \left( \matrix{ \delta_{L,R}^+/\sqrt{2} & \delta_{L,R}^{++} \cr
                          \delta_{L,R}^0 & -\delta_{L,R}^+/\sqrt{2} \cr }
              \right).
\end{equation}
The neutral Higgs fields $\delta_{R,L}^0\;,\;\phi_{1,2}^0$ can acquire vacuum
expectation values
\begin{equation}
<\Phi>=\left( \matrix{ k_1/\sqrt{2} & 0 \cr
                    0 & k_2/\sqrt{2} } \right) \;\;\;,\;\;\;
<\Delta_{L,R} > = \left( \matrix{ 0 & 0 \cr
                      v_{L,R}/\sqrt{2} & 0 } \right).
\end{equation}
{}From the present experimentally obtained bounds the additional gauge bosons
$W_2$ and
$Z_2$ are much heavier than the standard one $W_1,Z_1$ \cite{zralek}, so
$$v_R \gg k_1,k_2 \gg v_L,$$
and we have
\begin{eqnarray}
M_{W_1}^2&\simeq&\frac{g^2}{4} \left( k_1^2+k_2^2 \right) ,\\
M_{W_2}^2&\simeq& \frac{g^2}{2} v_R^2.
\end{eqnarray}
  The Yukawa interaction has the form
\begin{eqnarray}
L_{Yukawa}&\equiv&L_Y^B+L_Y^L+L_Y^R=-\bar{\Psi}_L \left[ h\phi + \tilde{h}
\tilde{\phi} \right] \Psi_R \nonumber \\
&-& \bar{\Psi}_LCi\tau_2 h_L\Delta_L\Psi_L
- \bar{\Psi}_RCi\tau_2h_R\Delta_R\Psi_R
\end{eqnarray}
where
$$\tilde{\phi}=\tau_2 \phi^{\ast} \tau_2,$$
and from left-right symmetry (1) we get
\begin{equation}
h=h^{\dagger}\;\;\;,\;\;\;\tilde{h}=\tilde{h}^{\dagger}\;\;\;,\;\;\;h_L=h_R.
\end{equation}
To find the physical Higgs bosons their interaction potential must be
specified.
We consider the most general Higgs potential which was discussed in Ref.
\cite{2intr}. To avoid the fine tuning problem the $\beta$ terms (Eq.(A2) in
\cite{2intr}) are made to vanish, $\beta_i=0$ (i=1,2,3). Then also $v_L$
has to vanish and $v_L=0$. The $\alpha_2$ parameter is assumed to
be real so there is no explicit CP violation (in the Higgs potential) and
the spontaneous CP symmetry breaking also does not appear (vacum expectation
values are real). Then the most general Higgs potential has the form
\cite{2intr}
\begin{eqnarray}
V=&-&\mu_1^2\left(Tr\left[\phi^{\dagger}\phi \right] \right)
- \mu_2^2\left(Tr\left[\tilde{\phi}\phi^{\dagger} \right] +
Tr\left[\tilde{\phi}^{\dagger} \phi \right]  \right) \nonumber \\
&-&\mu_3^2 \left( Tr \left[ \Delta_L \Delta_L^{\dagger} \right] +
 Tr \left[ \Delta_R \Delta_R^{\dagger} \right] \right) \nonumber \\
&+&\lambda_1 \left( \left( Tr \left[ \phi \phi^{\dagger} \right] \right)^2
\right)+ \lambda_2 \left( \left( Tr \left[ \tilde{\phi} \phi^{\dagger} \right]
  \right)^2 + \left( Tr \left[ \tilde{\phi}^{\dagger} \phi \right]
  \right)^2 \right) \nonumber \\
&+&   \lambda_3 \left(  Tr \left[ \tilde{\phi} \phi^{\dagger} \right]
   Tr \left[ \tilde{\phi}^{\dagger} \phi \right]  \right) \nonumber \\
&+& \lambda_4 \left( Tr \left[ \phi \phi^{\dagger} \right] \left(
   Tr \left[ \tilde{\phi} \phi^{\dagger} \right]
   + Tr \left[ \tilde{\phi}^{\dagger} \phi \right] \right) \right) \nonumber \\
&+& \rho_1 \left( \left( Tr \left[ \Delta_L \Delta_L^{\dagger} \right]
\right)^2+\left( Tr \left[ \Delta_R \Delta_R^{\dagger} \right]
\right)^2 \right) \nonumber \\
&+&\rho_2 \left( Tr \left[ \Delta_L \Delta_L \right]
Tr \left[ \Delta_L^{\dagger} \Delta_L^{\dagger} \right] +
Tr \left[ \Delta_R \Delta_R \right]
Tr \left[ \Delta_R^{\dagger} \Delta_R^{\dagger} \right]  \right) \nonumber \\
&+& \rho_3 \left(  Tr \left[ \Delta_L \Delta_L^{\dagger} \right]
Tr \left[ \Delta_R \Delta_R^{\dagger} \right] \right) \nonumber \\
&+&\rho_4 \left(  Tr \left[ \Delta_L \Delta_L \right]
Tr \left[ \Delta_R^{\dagger} \Delta_R^{\dagger} \right] +
Tr \left[ \Delta_L^{\dagger} \Delta_L^{\dagger} \right]
Tr \left[ \Delta_R \Delta_R \right] \right) \nonumber \\
&+& \alpha_1 \left( Tr \left[ \phi \phi^{\dagger} \right] \left(
Tr \left[ \Delta_L \Delta_L^{\dagger} \right] +
Tr \left[ \Delta_R \Delta_R^{\dagger} \right] \right) \right) \nonumber \\
&+&\alpha_2 \left( Tr \left[ \phi \tilde{\phi}^{\dagger} \right]
Tr \left[ \Delta_R \Delta_R^{\dagger} \right]+
Tr \left[ \phi^{\dagger} \tilde{\phi} \right]
Tr \left[ \Delta_L \Delta_L^{\dagger} \right] \right) \nonumber \\
&+&\alpha_2 \left( Tr \left[ \phi^{\dagger} \tilde{\phi} \right]
Tr \left[ \Delta_R \Delta_R^{\dagger} \right]+
Tr \left[ \tilde{\phi}^{\dagger} \phi \right]
Tr \left[ \Delta_L \Delta_L^{\dagger} \right] \right) \nonumber \\
&+& \alpha_3 \left( Tr \left[ \phi \phi^{\dagger} \Delta_L \Delta_L^{\dagger}
\right] +Tr \left[ \phi^{\dagger} \phi \Delta_R \Delta_R^{\dagger}
\right] \right).
\end{eqnarray}
The precise form of the mass matrices are given in Ref. \cite{2intr}.
 The 20 degrees of freedom in the Higgs sector give two charged
$G_{L,R}^{\pm}$,
 two neutral $G_{1,2}^0$ Goldston bosons and 14 physical degrees of freedom.
 These physical degrees of freedom produce:
 \begin{itemize}
 \item four neutral scalars with $J^{PC}=0^{++}$ ($H_i^0\;\;i=0,1,2,3$).
 \end{itemize}

 \
 Masses of three of them $H_0^0,H_1^0$ and $H_2^0$ are given from the
 diagonalization of the mass matrix which in the basis $ \left( \frac{\sqrt{2}}
 {y}Re\left( k_1\phi_1^0+k_2{\phi_2^0}^{\ast} \right), \right.$ \newline
 $\left. \frac{\sqrt{2}} {y}Re\left( k_1{\phi_2^0}^{\ast}-k_2\phi_1^0 \right),
 \sqrt{2}Re\delta_R^0
 \right) \equiv \left( \phi_-^{0r},\phi_+^{0r},\delta_R^{0r} \right) $
 has elements
\begin{eqnarray}
M_{11}&=& 2y^2 \left[\lambda_1+\epsilon^2 \left( 2\lambda_2+\lambda_3 \right)
+2\lambda_4\epsilon \right],  \nonumber \\
M_{12}=M_{21}&=& 2y^2\sqrt{1-\epsilon^2}  \left( 2\lambda_2+\lambda_3
+\lambda_4 \right) , \nonumber \\
M_{13}=M_{31}&=& \frac{1}{2}yv_R \left[2\alpha_1+2\alpha_2\epsilon+\alpha_3
\left(1-\sqrt{1-\epsilon^2} \right) \right], \nonumber \\
M_{22}&=&2\left( 2\lambda_2+\lambda_3 \right)y^2\left( 1-\epsilon^2 \right)+
\frac{1}{2}\alpha_3v_R^2\frac{1}{\sqrt{1-\epsilon^2}}, \\
M_{23}=M_{32}&=& \frac{1}{2}yv_R \left[ 4\alpha_2 \sqrt{1-\epsilon^2}+\alpha_3
\epsilon \right] ,\nonumber \\
M_{33}&=& 2\rho_1v_R^2, \nonumber
\end{eqnarray}
where
$$y^2=k_1^2+k_2^2 \;\;\;\mbox{\rm
and}\;\;\;\epsilon=\frac{2k_1k_2}{k_1^2+k_2^2}\;;
\;0\leq \epsilon \leq 1. $$
Then after diagonalization of the mass matrix we have
\begin{eqnarray}
\left( \matrix{\phi_-^{0r} \cr
               \phi_+^{0r} \cr
           \delta_R^{0r} } \right) =
\left( \matrix{ a_0 & a_1 & a_2 \cr
                b_0 & b_1 & b_2 \cr
        c_0 & c_1 & c_2 } \right)
\left( \matrix{ H_0^0   \cr
                H_1^0   \cr
        H_2^0 } \right),
\end{eqnarray}
where the parameters $a_i,b_i,c_i$ depend on the mass matrix elements (9).
If, however, $v_R \gg y$ we can find approximately that $H_0^0 \simeq
\phi_-^{0r} \;,\;H_1^0 \simeq \phi_+^{0r} \;,\; H_2^0 \simeq \delta_R^{0r}$
with the masses
\begin{eqnarray}
H^0_0\;, \hspace{2cm} M_{H_0^0}^2&\simeq & 2y^2\left[\lambda_1+\epsilon^2
\left( 2 \lambda_2+\lambda_3
\right) +2 \lambda_4\epsilon \right]  , \nonumber \\
H_1^0\;,\hspace{2cm} M^2_{H_1^0}& \simeq& \frac{1}{2}
\alpha_3v_R^2\frac{1}{\sqrt{1-\epsilon^2}}, \\
H_2^0\;,\hspace{2cm} M^2_{H_2^0}& \simeq& 2 \rho_1v_R^2. \nonumber
\end{eqnarray}
The lightest Higgs $H_0^0$ does not couple in neutral standard current to the
different flavour quarks (no FCNC). The other Higgs particles are heavy and the
FCNC, in the energy range which we consider, are reduced by the propagator
effect.
The fourth scalar Higgs particle $H_3^0$ has the mass
\begin{eqnarray}
H_3^0\;,\hspace{2cm} M_{H_3^0}^2 = \frac{1}{2} v_R^2\left( \rho_3-2\rho_1
\right) .
\end{eqnarray}
\begin{itemize}
\item two neutral pseudoscalars with $J^{PC}=0^{+-}$
\end{itemize}
\begin{eqnarray}
 A_1^0 ; \hspace{3cm}
 M_{A_1^0}^2 & =& \frac{1}{2}\alpha_3v_R^2\frac{1}{\sqrt{1-\epsilon^2}}-
 2y^2\left( 2\lambda_2-\lambda_3 \right), \nonumber \\
 A_2^0 ; \hspace{3cm} M_{A_2^0}^2 & = & \frac{1}{2} v_R^2 \left( \rho_3 -
 2 \rho_1 \right) .
\end{eqnarray}
\begin{itemize}
\item two singly charged bosons
\end{itemize}
\begin{eqnarray}
H_1^{\pm};\hspace{2cm}M_{H_1^{\pm}}^2 & =&\frac{1}{2}v_R^2\left( \rho_3-2\rho_1
\right) +
\frac{1}{4}\alpha_3y^2\sqrt{1-\epsilon^2}, \nonumber \\
H_2^{\pm};\hspace{2cm} M_{H_2^{\pm}}^2 & =&\frac{1}{2}\alpha_3 \left[v_R^2
\frac{1}{\sqrt{1-\epsilon^2}}+\frac{1}{2}y^2\sqrt{1-\epsilon^2} \right],
\end{eqnarray}
and finally
\begin{itemize}
\item two doubly charged Higgs particles
\end{itemize}
\begin{eqnarray}
\delta_L^{\pm\pm}; \hspace{3cm} M_{\delta_L^{\pm\pm}}^2 & =&
\frac{1}{2}\left[ v_R^2\left( \rho_3-2\rho_1 \right)+\alpha_3y^2
\sqrt{1-\epsilon^2} \right], \nonumber \\
\delta_R^{\pm\pm}; \hspace{3cm} M_{\delta_R^{\pm\pm}}^2 & =&
2\rho_2v_R^2+\frac{1}{2}\alpha_3y^2\sqrt{1-\epsilon^2}.
\end{eqnarray}
The parameters $\rho,\lambda,\alpha $ are defined in the Higss potential
(Eq.8).
The relations between the nonphysical Higgs particles specified in Yukawa
Lagrangian (Eq.6) and the physical ones are given by $\left(
k_{\pm}=\sqrt{k_1^2\pm
k_2^2}\right)$
\begin{eqnarray}
\phi^1_0&=&\frac{1}{\sqrt{2}k_+}\left[ H_0^0 \left( k_1 a_0-k_2b_0 \right)
+H_1^0 \left( k_1a_1-k_2b_1 \right) + H_2^0 \left( k_1a_2-k_2b_2 \right)
\right.
\nonumber \\
&+& \left. ik_1G_1^0-ik_2A_1^0 \right], \\
\phi^2_0 & = & \frac{1}{\sqrt{2}k_+}\left[ H_0^0 \left( k_2 a_0+k_1b_0 \right)
+H_1^0 \left( k_2a_1+k_1b_1 \right) + H_2^0 \left( k_2a_2+k_1b_2 \right)
\right.
\nonumber \\ &-& \left. ik_2G_1^0-ik_1A_1^0 \right], \\
\delta_L^0 &=& \frac{1}{\sqrt{2}} \left( H_3^0+iA_2^0 \right) ,\\
\delta_R^0 &=& \frac{1}{\sqrt{2}} \left( c_0H_0^0+c_1H_1^0+c_2H_2^0+iG_2^0
\right), \\
\phi_1^+&=&\frac{k_1}{k_+\sqrt{1+{\left( \frac{k_-^2}{\sqrt{2}k_+v_R}
\right)}^2}}H_2^+ -\frac{k_1}{k_+\sqrt{1+{\left( \frac{\sqrt{2}k_+v_R}{k_-^2}
\right)}^2}}G_R^+  - \frac{k_2}{k_+}G_L^+ ,\\
\phi_2^+&=&\frac{k_2}{k_+\sqrt{1+{\left( \frac{k_-^2}{\sqrt{2}k_+v_R}
\right)}^2}}H_2^+ -\frac{k_2}{k_+\sqrt{1+{\left( \frac{\sqrt{2}k_+v_R}{k_-^2}
\right)}^2}}G_R^+ + \frac{k_1}{k_+}G_L^+ ,\\
\delta_L^+ &=& H_1^+ ,\\
\delta_R^+&=&
\frac{1}{\sqrt{1+{\left( \frac{k_-^2}{\sqrt{2}k_+v_R}
\right)}^2}}G_R^+  + \frac{1}{\sqrt{1+{\left( \frac{\sqrt{2} k_+v_R}{k_-^2}
\right)}^2}}H_2^+ .
\end{eqnarray}
If $v_R \gg y$ then the relations for $\phi_{1,2}^0,\phi_{1,2}^+,\delta_R^0$
and $\delta_R^+$ are simpler
\begin{eqnarray}
\phi_1^0&\simeq&\frac{1}{y\sqrt{2}}\left[ k_1H_0^0-k_2H_1^0+ik_1G_1^0-ik_2A_1^0
\right] ,\\
\phi_2^0 &\simeq& \frac{1}{y\sqrt{2}}\left[ k_2H_0^0+k_1H_1^0-ik_2G_1^0-
ik_1A_1^0 \right], \\
\delta_R^0 &=& \frac{1}{\sqrt{2}}\left( H_2^0+iG_2^0 \right) ,\\
\phi_{1,2}^+&\simeq& \left[ \frac{1\pm\sqrt{1-\epsilon^2}}{2}
\right]^{1/2}H_2^+
-\frac{y\left[ 1-\epsilon^2 \pm \sqrt{1-\epsilon^2}\right]^{1/2}}{2v_R}G_R^+
\nonumber \\
&-&\left[ \frac{1\mp\sqrt{1-\epsilon^2}}{2}\right]^{1/2}G_L^+, \\
\delta_R^+ &\simeq& G_R^++\frac{y\sqrt{1-\epsilon^2}}{\sqrt{2}v_R}H_2^+,
\end{eqnarray}
where
$$k_1=y\left[ \frac{1+\sqrt{1-\epsilon^2}}{2} \right]^{1/2}\,;\,k_2=y
\left[ \frac{1-\sqrt{1-\epsilon^2}}{2} \right]^{1/2}.
$$
The doubly charged Higgs bosons are already physical. $G_{1,2}^0$ and $G_{L,R}^
{\pm}$ Goldston bosons are `eaten' by $Z_{1,2}$ and $W_{L,R}^{\pm}$
respectively.

\ \ \ Now we can express the Yukawa interaction Lagrangian in the terms of
physical
quantities. For the bidoublet interaction
\begin{eqnarray}
-L_Y^B&=&
\bar{\nu}_L \left( h\phi_1^0+\tilde{h}{\phi_2^0}^{\ast} \right) \nu_R +
\bar{\nu}_L \left( h\phi_1^+ -\tilde{h}\phi_2^+ \right) e_R \nonumber \\
&+&e_L \left( h\phi_2^- -\tilde{h}\phi_1^- \right) \nu_R+
e_L \left( h{\phi_2^0}+\tilde{h}{\phi_1^0}^{\ast} \right) e_R + h.c.,
\end{eqnarray}
one has
\begin{eqnarray}
&&\bar{\nu}_L \left( h\phi_1^0+\tilde{h}{\phi_2^0}^{\ast} \right) \nu_R + h.c.=
\hspace{5cm}
\nonumber \\
& \sum_{a} & \bar{N}_a \left\{ \left[ \left( \Omega_L \right)_{aa}m^N_a
B_0 + \sum_{l} \left( K_L \right)_{al} \left( K_R^{\ast}
\right)_{al} m_lA_0^{\ast} \right] P_R  \right. \nonumber \\
&+& \left. \left[ m^N_a \left(
\Omega_L \right)_{aa} B_0^{\ast} + \sum_{l} \left( K_L^{\ast} \right)_{al}
\left(
 K_R \right)_{al} m_lA_0 \right] P_L \right\} N_a \nonumber \\
&+&\sum_{a>b} \bar{N}_a \left\{ \left[ \left( \left( \Omega_L \right)_{ac}m^N_c
\left( \Omega_R \right)_{cb} + \left( \Omega_L \right)_{bc}m^N_c
\left( \Omega_R \right)_{ca} \right) B_0 \right. \right. \nonumber \\
&+& \left.  \sum_{l} m_l \left( \left( K_L
\right)_{al} \left( K_R^{\ast} \right)_{bl} + \left( K_L
\right)_{bl} \left( K_R^{\ast} \right)_{al} \right) A_0^{\ast} \right] P_R
\nonumber \\
\nonumber \\
&+& \left[ \left( \left( \Omega_L \right)_{ac}m^N_c
\left( \Omega_R \right)_{cb} + \left( \Omega_L \right)_{bc}m^N_c
\left( \Omega_R \right)_{ca} \right) B_0^{\ast} \right. \nonumber \\
&+& \left. \left.  \sum_{l} m_l \left( \left(
K_L^{\ast} \right)_{bl} \left( K_R \right)_{al} + \left( K_L^{\ast}
\right)_{al} \left( K_R \right)_{bl} \right) A_0 \right] P_L
\right\} N_b, \\
&& \nonumber \\
&&\bar{\nu}_L \left( h\phi_1^+ -\tilde{h}\phi_2^+ \right) e_R +
\bar{e}_L \left( h\phi_2^- -\tilde{h}\phi_1^- \right) \nu_R + h.c. = \nonumber
\\
&+&\sum_{a,l}\bar{N}_a \left\{ \left[ \sum_{b} \left( \Omega_L \right)_{ab}
m_b^N \left( K_R \right)_{bl} A^+ -\left( K_L \right)_{al} m_l
B^+ \right] P_R \right. \nonumber \\
&+& \left. \left[ m^N_a \left( K_L \right)_{al}
B^+ -\left( K_R \right)_{al}m_l A^+ \right] P_L \right\} e_l \nonumber \\
&+&\sum_{a,l}\bar{e}_l \left\{ \left[ \sum_{b} \left( K_R^{\dagger}
\right)_{lb}
m_b^N\left( \Omega_L \right)_{ba} A^- -m_l \left( K_L^{\dagger} \right)_{la}
B^- \right] P_L \right. \nonumber \\
&+& \left. \left[ \left( K_L^{\dagger} \right)_{la} m_a^N
B^- - m_l \left( K_R^{\dagger} \right)_{la} A^- \right] P_R \right\} N_a ,
\end{eqnarray}
and
\begin{eqnarray}
&&\bar{e}_L \left( h{\phi_2^0}+\tilde{h}{\phi_1^0}^{\ast} \right) e_R + h.c.=
\nonumber \\
& \sum_{l,k}& \bar{e}_l \left\{ \left[ \delta_{lk} m_l B_0^{\ast} +
\sum_{a} \left( K_L^{\ast} \right)_{al} \left( K_R \right)_{ak} m_a^N A_0
\right] P_R \right. \nonumber \\
&+& \left. \left[ \delta_{lk} m_l B_0 +
\sum_{a} \left( K_L \right)_{ak} \left( K_R^{\ast} \right)_{al} m_a^N
A_0^{\ast}
\right] P_L \right\} e_k.
\end{eqnarray}
$m_b^N$ denote neutrino masses, three light (b=1,2,3) and three heavy
(b=4,5,6).
$m_l$ are the charged lepton masses (l=$e,\mu,\tau$).
For the definition of matrices $K_{L,R}$ and $\Omega_{L,R}$ see Ref.
\cite{4intr}.
The parameters $A_0,B_0,A^{\pm}$ and $B^{\pm}$ denote the combination of the
Higgs fields
\begin{eqnarray}
A_0&=&\frac{\sqrt{2}}{k_-^2}\left(k_1\phi_2^0-k_2{\phi_1^0}^{\ast} \right)
,\nonumber \\
B_0&=&\frac{\sqrt{2}}{k_-^2}\left(k_1\phi_1^0-k_2{\phi_2^0}^{\ast} \right) ,
\nonumber \\
A^{\pm}&=&\frac{\sqrt{2}}{k_-^2}\left(k_1\phi_1^{\pm} +k_2{\phi_2^{\pm}}
\right) , \nonumber \\
B^{\pm}&=&\frac{\sqrt{2}}{k_-^2}\left(k_2\phi_1^{\pm} +k_1{\phi_2^{\pm}}
\right).
\end{eqnarray}
The right-handed triplet interaction with leptons is
\begin{eqnarray}
-L_Y^R&=&\delta_R^0\bar{\nu}_L^ch_R\nu_R-
\frac{\delta_R^+}{\sqrt{2}}\left( \bar{\nu}_L^c h_Re_R+\bar{e}_L^ch_R\nu_R
\right) \nonumber \\
&-&\delta_R^{++}\bar{e}_L^ch_Re_R+h.c.,
\end{eqnarray}
where the first two parts are given by
\begin{eqnarray}
&&\delta_R^0\bar{\nu}_L^ch_R\nu_R + h.c.= \nonumber \\
&-&\frac{1}{2v_R} \left\{ \sum_{a} \bar{N}_a \left( \sum_{c} m_c^N \left[
\left( \Omega_R \right)^2_{ca} P_R+ \left( \Omega_R^{\ast} \right)^2_{ca} P_L
\right] \right) N_a \right. \nonumber \\
&-& \left. \frac{1}{v_R} \sum_{a>b} \bar{N}_a \left( \sum_{c} m_c^N \left[
\left( \Omega_R \right)_{cb} \left( \Omega_R \right)_{ca}P_R+
\left( \Omega_R^{\ast} \right)_{cb} \left( \Omega_R^{\ast} \right)_{ca} P_L
\right] \right) N_b \right\}(\sqrt{2}Re \delta^0_R) \nonumber \\
&+&\mbox{\rm  Goldston boson interaction,}
\end{eqnarray}
and
\begin{eqnarray}
&&-\frac{\delta_R^+}{\sqrt{2}}\left( \bar{\nu}_L^c h_Re_R+\bar{e}_L^ch_R\nu_R
\right) + h.c. = \nonumber \\
&-&\frac{1}{v_R} \sum_{a,l} \left\{ \bar{N}_a \left[ \sum_{b} \left( \Omega_R^
{\ast} \right)_{ab} m_b^N \left( K_R \right)_{bl} \right] P_Re_l \delta_R^+
\right. \nonumber \\
&+& \left. \bar{e}_l \left[ \sum_{b} \left( K_R^{\dagger} \right)_{lb} m_b^N
\left( \Omega_R^{\ast} \right)_{ba} \right] P_LN_a \delta_R^- \right\}.
\end{eqnarray}
To get the interaction of the left-handed triplet the indices L(R) in the
formulae
(35) and (36) should be replaced by R(L) (if $v_L\neq 0$). If the left handed
triplet does not condensate ($v_L=0$) it will still interact with the leptons
($h_L=h_R\neq 0$).
Then to get its interaction with physical leptons we have to use
the formulae
\begin{eqnarray}
&&\delta_L^0 \bar{\nu}_R^ch_L\nu_L + h.c. = \nonumber \\
& \frac{1}{\sqrt{2}v_R} & \sum_{a} \bar{N}_a \left[ X_{aa}\delta_L^0P_L
+X_{aa}^{\ast}{\delta_L^0}^{\ast}P_R \right] N_a \nonumber \\
&+&  \frac{\sqrt{2}}{v_R} \sum_{a>b} \bar{N}_a \left[
X_{ab}\delta_L^0P_L+X_{ab}^{\ast}{\delta_L^0}^{\ast}P_R \right] N_b ,
\end{eqnarray}
and
\begin{eqnarray}
&&-\frac{\delta_L^+}{\sqrt{2}}\left(\bar{\nu}_R^c h_Le_L+\bar{e}_R^ch_L\nu_L
\right) + h.c. = \\
&-&\frac{1}{v_R} \delta_L^+ \sum_{a,l} \bar{N}_a \sum_{b}X_{ab}
\left( K_L \right)_{bl}P_Le_l +\frac{1}{v_R}
\delta_L^- \sum_{a,l} \bar{e}_l \sum_{b} \left( K_L^{\dagger} \right)_{lb}
\left( X^{\ast}\right)_{ba} P_RN_a, \nonumber
\end{eqnarray}
where the matrix X is given by
\begin{equation}
X_{ab}=\sum_{c}\left( U_R^{\dagger}U_L \right)^T_{ac} m_c^N \left(
U_R^{\dagger}
U_L \right)_{cb}=X_{ba}.
\end{equation}
\section{Parametrization of mass and mixing matrices for leptons.}
\ \ \
In the model which we consider the left-handed triplet does not acquire the
vacuum expectation value $\left( v_L=0 \right) $ so the neutrino mass matrix
has  the form
\begin{equation}
M^{\nu}=\left( \matrix{ 0 & M_D \cr
                  M_D^T & M_R } \right) ,
\end{equation}
where
\begin{equation}
M_D=\frac{1}{\sqrt{2}}\left( hk_1+\tilde{h}k_2 \right)\;\; \mbox{\rm and}\;\;
M_R=\sqrt{2}h_Rv_R
\end{equation}
are hermitian and symmetric $3 \times 3$ matrices respectively.
The charged leptons' masses come from the same $h$ and $\tilde{h}$ terms
\begin{equation}
M_l=\frac{1}{\sqrt{2}}\left( hk_2+\tilde{h}k_1 \right).
\end{equation}
We can perform a unitary transformation, the same for the left and the
right-handed lepton fields (L-R symmetry), without changing the physical
interpretation of the theory. It means that the matrices
\begin{eqnarray}
M_D & \leftrightarrow & VM_DV^{\dagger} ,\nonumber \\
M_R & \leftrightarrow & V^{\ast} M_R V^{\dagger}, \\
M_l & \leftrightarrow & VM_lV^{\dagger} \nonumber
\end{eqnarray}
are totally equivalent from the physical point of view \cite{sant} .
Using the equivalence relations (Eq.43) we can diagonalize the $M_R$ matrix
leaving
the matrices $M_D$ and $M_l$ still hermitian. So we see that the lepton sector
of our theory with the left-right symmetry (Eq.1) is described by 6+6+3=15
moduli (9 masses + 6 angles) and 3+3=6 CP violating phases.
\begin{equation}
M_l= \left( \matrix{ a_l & b_le^{i\beta_l} &c_le^{i\gamma_l} \cr
                     b_le^{-i\beta_l} & d_l & f_le^{i\delta_l} \cr
             c_le^{-i\gamma_l} & f_le^{-i\delta_l} & g_l} \right) \;\;,\;\;
M_D= \left( \matrix{ a_D & b_De^{i\beta_D} &c_De^{i\gamma_D} \cr
                     b_De^{-i\beta_D} & d_D & f_De^{i\delta_D} \cr
             c_De^{-i\gamma_D} & f_De^{-i\delta_D} & g_D} \right) ,
\end{equation}
and
\begin{equation}
M_R=diag(\bar{M}_1,\bar{M}_2,\bar{M_3}).
\end{equation}
The neutrino mass matrix $M^{\nu}$ and the charged lepton mass matrix $M_l$
are diagonalized
by the orthogonal and unitary transformations respectively
\begin{eqnarray}
U^TM^{\nu}U&=&diag(m_1,m_2,m_3,M_1,M_2,M_3)\equiv M^{\nu}_{diag} \nonumber \\
U^{\dagger}_lM_lU_l&=&diag(m_e,m_{\mu},m_{\tau})\equiv m^l_{diag}.
\end{eqnarray}
We use the procedure (see e.g. Ref. \cite{buchpil}) that gives us means to find
the matrix U in the approximate way. As $v_R \gg {\left[ k_1^2+k_2^2
\right]}^{1/2}$
we assume that
\begin{equation}
\bar{M}_i \gg a_D,b_D,\ldots ,g_D\;\;\;\;\;\;\;(i=1,2,3)
\end{equation}
so the elements of the matrix
\begin{equation}
\xi=M_DM_R^{-1}
\end{equation}
are small
$$ \mid \xi_{ij} \mid \ll 1. $$
Then the form of matrix U to the third order in $\xi$ can be estimated as
\begin{equation}
U \equiv \left( \matrix{ U_L^{\ast} \cr
                U_R   } \right) =
\left( \matrix{ \left[1-\frac{1}{2}\xi^{\ast}\xi^T \right] J_l &
               \left[ \xi^{\ast} \left( 1-\frac{1}{2}\xi^T \xi^{\ast} \right)
           \right] J_h \cr
           \left[ -\xi^T \left( 1-\frac{1}{2}\xi^{\ast}\xi^T \right)
           \right] J_l &
           \left[  1-\frac{1}{2}\xi^T \xi^{\ast} \right] J_h }
           \right) ,
\end{equation}
where $J_l$ and $J_h$ are diagonal unitary matrices that guarantee that the
mass eigenvalues are positive. The diagonal elements of matrices $J_l$ and
$J_h$
are equal 1 or i and in case of CP conservation describe the CP parity
$\eta_{CP}$
of the appropriate Majorana neutrino
\begin{eqnarray*}
\eta_{CP}&=&+i\;\;\; \mbox{\rm    if the diagonal element equals +1}   \\
\eta_{CP}&=&-i\;\;\; \mbox{\rm    if the diagonal element equals +i.}
\end{eqnarray*}
Now we can easily find the all necessary mixing matrices
\begin{eqnarray}
&&K_L=U_L^{\dagger}U_l  =  \left( \matrix{ J_l \left( 1 -\frac{1}{2}\xi\xi^
{\dagger} \right) U_l \cr
J_h \left( 1 -\frac{1}{2}\xi\xi^{\dagger} \right) \xi^{\dagger} U_l }
\right)\;, \\
&& \nonumber \\
&&K_R=U_R^{\dagger}U_l  = \left( \matrix{ -J_l^{\ast} \left( 1 -\frac{1}{2}
\xi^{\ast}\xi^T \right) \xi^{\ast} U_l \cr
J_h^{\ast} \left( 1 -\frac{1}{2}
\xi^T\xi^{\ast} \right)  U_l } \right) \;, \\
&& \nonumber \\
&&\Omega_L=U_L^{\dagger}U_L= \left( \matrix{ J_l \left( 1 -\xi\xi^
{\dagger} \right) J_l^{\ast} & J_l \xi \left( 1 -\xi^
{\dagger}\xi \right) J_h^{\ast} \cr
J_h \xi^{\dagger} \left( 1 -\xi\xi^
{\dagger} \right) J_l^{\ast} & J_h \xi^{\dagger}\xi J_h^{\ast} } \right) ,\; \\
&& \nonumber \\
&&\Omega_R=U_R^{\dagger}U_R= \left( \matrix{ J_l^{\ast} \xi^{\ast}\xi^T
J_l & -J_l^{\ast} \xi^{\ast} \left( 1 -\xi^T\xi^
{\ast} \right) J_h \cr
-J_h^{\ast} \xi^T \left( 1 -\xi^{\ast}\xi^T \right) J_l & J_h^{\ast} \left(1-
\xi^T\xi^{\ast} \right) J_h } \right), \;
\end{eqnarray}
and
\begin{eqnarray}
\Omega_{RL}&\equiv &U_R^{\dagger}U_L= \\
&& \nonumber \\
&&\left( \matrix{ -J_l^{\ast}\xi^{\ast}
\left( 1-\frac{1}{2}\xi^T\xi^{\ast}-\frac{1}{2}\xi\xi^{\dagger} \right)
J_l^{\ast} & -J_l^{\ast} \xi^{\ast}\xi J_h^{\ast} \cr
J_h^{\ast}
\left( 1-\frac{1}{2}\xi^T\xi^{\ast}-\frac{1}{2}\xi\xi^{\dagger} \right) J_l^
{\ast} &
J_h^{\ast} \left( 1-\frac{1}{2}\xi^T\xi^{\ast}-\frac{1}{2}\xi\xi^{\dagger}
\right) \xi J_h^{\ast} } \right) .\nonumber
\end{eqnarray}
If the CP symmetry holds then the matrices $h$,$\tilde{h}$ and $h_L=h_R$
are real and the phases in matrices $M_D$ and $M_l$ disappear. If there are
neutrinos with opposite CP parities then certain columns in the matrices K and
the relevant rows and columns in matrices $\Omega$ are pure imaginary.
\section{Higgs particles influence on heavy Majorana neutrino production:
numerical results.}
\ \ \
The full cross section for the $e^-e^+ \rightarrow \nu N$ process is described
in
our model by six diagrams with $W_{1,2}$ (t and u channels) and $Z_{1,2}$
(s channel) exchange
(Fig.1) and by the Higgs exchange diagrams in all three channels (Fig.2).
The influence of the spin and boson exchange diagrams are discussed in details
in
Ref.\cite{4intr}. Here we would like to ascertain about the role of Higgs
exchange diagrams.
In the t and u channels two charged Higgs particles can be exchanged
(Fig.2a,b).
In the approximation which we consider $\left( v_R \gg y\;,\;m_e\simeq0
\right)$
only two neutral Higgses couple in the s channel. All the calculations are
done in the unitary gauge, so we do not take into account the Goldston
particles
exchange.
The full helicity amplitude for the process $e^-e^+ \rightarrow \nu N$ and the
precise
values of all couplings are presented in the Appendix.
We can see
that like in the SM the lighter Higgs particle $H_0^0$ couple to the
$e^-e^+$ proportionally to the electron mass and its efect is negligible in the
energy range which we consider. The influence of two charged $H_{1,2}^+$ and
two
neutral $\left( H_1^0,A_1^0 \right)$ Higgs particles is not obvious. At first
sight
their coupling, even to the light leptons $\left( e^-e^+,e\nu\right), $ can be
large
as there are terms in the vertex proportional to heavy neutrinos mass. They
are,
however, multiplied by the mixing matrices which can have small terms so the
total efect needs numerical analysis.
As an example let's take the $M_D$ and $M_R$ mass matrices in the form
\begin{eqnarray}
M_D=\left( \matrix{ 1. & 1. & 0.9 \cr
                1. & 1. & 0.9 \cr
                0.9 & 0.9 & 0.95 } \right)\;\;,\;\;
M_R=\left( \matrix{ 10^2 & 0 & 0 \cr
                0 & 10^3 & 0 \cr
                0 & 0 & 10^6 } \right),
\end{eqnarray}
which produce the realistic spectrum of neutrino masses $\left( m_{\nu_e}
= 0\;,\right.$ \newline $\left.
m_{\nu_{\mu}}\simeq1.7eV\;,\;m_{\nu_{\tau}}=33MeV\;,\;
M_1=100GeV\;,\;M_2=10^3GeV\;,\right.$ \newline
$\left. \;M_3=10^6GeV \right).$
The matrix U which diagonalizes neutrino mass matrix $M^{\nu}$ and gives
positive
eigenvalues of masses can be calculated precisely and is equal
\newline
$$U=$$
\begin{eqnarray*}
\left( \matrix{ .707 & -.408 i & -.577 i & -.010 & -.001 & .9\cdot10^{-6} \cr
               -.707 & -.408 i & -.577 i & -.010 & -.001 & .9\cdot10^{-6} \cr
               .6\cdot10^{-12} & .816 i & -.577  i &-.010 & -.001 &
1.\cdot10^{-6} \cr
              -.1\cdot10^{-16} & .3\cdot10^{-7} i& .017 i &-1. &
-.3\cdot10^{-5} &
                                          .3\cdot10^{-11} \cr
            -.1\cdot10^{-16} & .3\cdot 10^{-8}  i& .002 i&.3\cdot10^{-4} & -1.
&
                                                 .3\cdot10^{-11} \cr
             -.3\cdot10^{-19} & -.4\cdot 10^{-7} i & .2\cdot 10^{-5} i&
               .3\cdot 10^{-7}  &             .3\cdot10^{-8} & 1. }
                                     \right)
\end{eqnarray*}

\
from which all other necessary mixing matrices (50)-(54) can be calculated.
To find the
role of the Higgs sector in our case we calculate the cross section for three
different processes $e^-e^+ \rightarrow \nu_eN(100)\;,\;\nu_{\mu}N(100)\;$
and $\nu_{\tau}N(100)$. The values of $v_R$ and y are obtained from massess
$M_{W_1}^2
\simeq \frac{g^2}{4}y^2$ and $ M_{W_2}^2\simeq \frac{g^2}{2}v_R^2$.
To calculate the Higgs boson masses we use the relations (11)-(15) with the
Higgs potential
parameters (or some their combinations) equal to 1, so we have
\begin{eqnarray}
M_{H_1^{\pm}}^2&=&\frac{1}{2} \left[ v_R^2 + \frac{1}{2}y^2\sqrt{1-\epsilon^2}
\right], \nonumber \\
M_{H_2^{\pm}}^2&=&\frac{1}{2} \left[ v_R^2\frac{1}{\sqrt{1-\epsilon^2}} +
\frac{1}{2}y^2\sqrt{1-\epsilon^2} \right], \nonumber \\
M_{H_1^{0}}^2&\simeq&\frac{1}{2}  v_R^2\frac{1}{\sqrt{1-\epsilon^2}}, \nonumber
\\
M_{A_1^{0}}^2&\simeq&\frac{1}{2} \left[ v_R^2\frac{1}{\sqrt{1-\epsilon^2}}
-4y^2 \right].
\end{eqnarray}
For $M_{W_1}\simeq 80 GeV$ and $M_{W_2}\simeq 1600 GeV$ (the value accepted
from
experimental data Ref.[9]) we find the range for y and $v_R$
\begin{eqnarray}
y & \simeq & 250 GeV, \nonumber \\
v_R &\simeq & 3500 GeV.
\end{eqnarray}
Then for $\epsilon =0$ all Higgs boson masses from Eq.(56) are of 2.5 TeV order
\begin{equation}
M_{H_1^{\pm}}\ \simeq M_{H_2^{\pm}} \simeq M_{H_1^0} \simeq M_{A_1^0} \simeq
2450 GeV.
\end{equation}
For the neutral Higgs bosons $H_1^0$ and $A_1^0$ which cause the FCNC the
masses
of order 2.5 TeV are not large enough to reduce the $\bar{K}^0-K^0$ transition.
To generate proper mass spliting in $\bar{K}^0-K^0$ system it was found that
Ref.[10]
\begin{equation}
M_{H_1^0}, M_{A_1^0} > 10 TeV.
\end{equation}
There are two ways of obtaining these large mass values for the FCNC neutral
Higgs bosons
\begin{itemize}
\item we can choose the parameter $\alpha_3$ in Eq.(11) and (13) to be greater
then 1 e.g. $\alpha_3 \simeq 16$. Then $M_{H_1^0}, M_{A_1^0} $ satisfy the
bound (59) and the charged bosons can be lighter $M_{H_1^{\pm}}\simeq M_{H_2^
{\pm}} \simeq 2500 GeV$ (for $\epsilon$ not too close to 1),\newline
or
\item
we avoid the fine tuning for the Higgs parameters ($\alpha_3 \simeq 1$) but we
assume that the $\phi_1^0$ and $\phi_2^0$ acquire approximately
the same VEV $k_1\simeq k_2$, so then $\epsilon \simeq 1$ and the masses
$M_{H_1^0}, M_{A_1^0} $ and $M_{H_2^{\pm}}$ are much greater then $\frac{1}{2}
v_R^2$ and can also satisfy the bound (59).
\end{itemize}
In the first case the presence of large masses in the propagator causes that
the total
contribution of the $H_1^0$ and $A_1^0$ exchange in the s channel is very
small.
In the second case, when $\epsilon \rightarrow 1$, the couplings of neutral
Higgses
$H_1^0$ and $A_1^0$ to leptons become stronger (see (A.18)) and compensate the
influence of the propagator. The total effect depends on the additional vertex
contributions given in (A.19) and (A.20). We calculate numerically the factors
of the type
\begin{equation}
(K_L^{\dagger}M^{\nu}_{diag}K_R)_{ee}\;\;,\;\;(K_Lm_l^{diag}K_R^{\dagger})_{ab}\;\;,\;\;
(\Omega_LM^{\nu}_{diag})_{ab}
\end{equation}
which are present in the couplings (A.19) and (A.20) for different values of
the
heavy neutrino mass. The factors are of the same order independently of the
neutrino masses.
It is caused by the fact that for bigger neutrino masses the appropriate
mixing matrix elements are smaller. We have calculated the influence of the
scalar $H_1^0$ and pseudoscalar $A_1^0$ exchange diagrams on the total cross
section.
For the energy range which we consider (200-500 GeV) this contribution is
completely
negligible. We have checked also the contribution of the charged Higgs bosons
exchange
in the t-u channels.
In Table 1 we present the ratios of cross sections with only gauge bosons
($\sigma_{gauge}$) or Higgs particles ($\sigma_{Higgs}$) to $\sigma_{total}$
in which all Feynman diagrams are taken into account. We can see that Higgses
have
no meaning for heavy Majorana neutrino production. For $\nu_{\mu}N$ and
$\nu_{\tau}N$ neutrino production the Higgs exchange mechanism gives only the
contribution of order $10^{-4}$.
\begin{table}
\centering
\begin{tabular}{c c c}
\cline{1-3}
  &\ &\ \\
$e^-e^+$ &\  $\sigma_{gauge}/\sigma_{total}$ &\ $\sigma_{Higgs}/\sigma_
{total}$ \\
 &\ &\ \\
\cline{1-3}
$\rightarrow \nu_eN(100)$ & $\simeq 100\%$ & $\simeq .0001\%$ \\
  &\ &\ \\
$\rightarrow \nu_{\mu}N(100)$ & $\simeq 100\%$ & $\simeq .01\%$ \\
  &\ &\ \\
$\rightarrow \nu_{\tau}N(100)$ & $\simeq 100\%$ & $\simeq .01\%$ \\
\cline{1-3}
\end{tabular}
\caption{The contribution of the gauge and Higgs bosons to the total cross
section
for LEPII energy.}
\end{table}
Moreover, these results are not sensitive to the $\epsilon$ factor. The Feynman
diagram
with $H_1^+$ exchange (which is the most important)  is not sensitive
to this factor at all (A.15).
 The $H_2^+$ exchange diagram is sensitive to this factor and the
contribution to the total cross section increases with increasing $\epsilon$,
but as the propagator for this particle is sensitive to this factor too,
the increase is rather small and even for $\epsilon=1$ does not predominante
the $H_1^+$ contribution. The result is shown in Fig.4. Finally, in Fig.5.
we gather all processes that give a single heavy neutrino production with
mass equal 100 GeV as the CM energy function. For LEPII it gives about 2 heavy
neutrino productions per year ($L/yr=500pb^{-1}$) and about 100 events per year
for NLC energy ($L/yr=10fb^{-1}$).
It is worth to mention that for smaller $M_{W_2}$ also $H_{1,2}^{\pm}$ Higgs
particles masses can be smaller and the total Higgs contribution will be
larger.
\section{Conclusions}
\ \ \
We have investigated in details the importance of the Higgs boson exchange
diagrams in
the neutrino production in $e^-e^+$ collisions. For the general Higgs potential
in the left-right symmetric model we have found the physical eigenmass states.
The couplings of these physical Higgs particles with leptons are expressed in
terms
of physical quantities (masses, mixing angles, CP violating phases). If we
neglect the electron's mass, only four non-standard Higgs bosons couple with
$e^-e^+$
and $\nu N$ systems- two charged $H_{1,2}^{\pm}$ and two neutral ones, one
scalar
$H_1^0$, and one pseudoscalar $A_1^0$. As the result of lack of the FCNC,
masses
of the neutral Higgs bosons must be large ($>$ 10 TeV). The influence of such
Higgs particles on the $e^-e^+ \rightarrow \nu N$ cross section is very small
even if their coupling with leptons is of the same order as the gauge bosons
couplings. The couplings of the charged Higgs bosons are also comparable to the
gauge bosons ones and their small contribution  $(10^{-4})$ to the cross
section is caused mainly by the bigger mass in the Higgs propagators.
The contribution of all Higgs exchange diagrams to the full
neutrino production $e^-e^+ \rightarrow \nu N$ cross section is small and
almost
independent of the masses of the heavier neutrinos.
\section*{Appendix}
\setcounter{equation}{0}
\renewcommand{\theequation}{A.\arabic{equation}}
\ \ \
We present here the full helicity amplitudes for the process
\begin{equation}
e^-(p,E;\sigma)+e^+(p,E;\bar{\sigma}) \rightarrow \nu (q,E';\lambda)
+ N(q,E'';\bar{\lambda}).
\end{equation}
 In brackets momentum,energy and helicities of particles are denoted.
The other symbols we use are
$$
\beta'=\frac{q}{E}\;,\;\beta''=\frac{q}{E''}\;,\;\Delta\sigma=\sigma-
\bar{\sigma}\;\;\mbox{\rm and}\;\;\Delta\lambda=\lambda-\bar{\lambda}.
$$
The gauge bosons couplings are precisely defined in Ref. \cite{4intr}.
\def\sig{\sigma\bar{\sigma}}
\def\lam{\lambda\bar{\lambda}}
\def\bds{\mid \Delta \sigma \mid}
\def\bdl{\mid \Delta \lambda \mid}
\begin{eqnarray}
&&-iM \left( \sig ; \lam \right) = i \frac{s}{2}\sqrt{1-{ \left(
\frac{M_N^2-M_{\nu}^2} {s} \right) }^2} \sum_{i=1}^{2} \Biggl\{
D_{\Delta\sigma,\Delta\lambda}^{1^{\ast}}\left(\phi,\Theta,0\right)  \nonumber
\\
&&  \nonumber \\
&&\Biggl(  {\left( \sqrt{2}  \right)}^{\bds +\bdl}
\left[ \frac{A_1^{t_i}}{t-M_{W_i}^2}-\frac{A_1^{u_i}}{u-M_{W_i}^2}+
\frac{A^{s_i}}{s-M_{Z_i}^2+i\Gamma_i M_{Z_i}} \right]\; +   \nonumber \\
&&                          \nonumber   \\
&&  {\left( \sqrt{2} \right)}^{-\mid \bds - \bdl \mid} \left[
\frac{A_2^{t_i}}{t-M_{W_i}^2} - \frac{A_2^{u_i}}{u-M_{W_i}^2}
+  \frac{A_{H_i}^t}{t-M_{H_i}^2}- \frac{A_{H_i}^u}{u-M_{H_i}^2} \right] \Biggr)
\nonumber \\
&&                               \nonumber           \\
&&+ {\delta}_{\sigma\bar{\sigma}}
\delta_{\lambda\bar{\lambda}} \biggl[
\frac{h^{t_i}+f^{t_i}}{t-M_{W_i}^2}-\frac{h^{u_i}+f^{u_i}}
{u-M_{W_i}^2} \nonumber \\
&&+ \frac{g^t_{H_i}}{t-M_{H_i}^2}-\frac{g^u_{H_i}}{u-M_{H_i}^2}
+ \frac{g^s_{H_i^0}}{s-M_{H_i^0}^2+i\Gamma_{H_i} M_{H_i^0}}
\biggr]  \Biggr\},
\end{eqnarray}
where the sum $\sum_{i=1}^{2}$ is over $W_{1,2}^{\pm}$,$Z_{1,2}^0$,
$H_{1,2}^{\pm}$ and $H_1^0$ or $A_1^0$. All other symbols are explained in
Ref. \cite{4intr}.
\begin{eqnarray}
A_1^{t_i}\left( \sig;\lam\right)&=&
{\left(A_L^{lN_i}\right)}^{\ast}_{Ne} {\left(A_L^{lN_i}\right)}_{\nu e}
{\delta}_{\Delta\sigma,-1} \sqrt{( 1-2\lambda\beta')
(1+2\bar{\lambda}\beta'') } \ \ \ \ \ \ \ \ \ \ \ \ \ \ \ \nonumber \\
&&+{\left(A_R^{lN_i}\right)}^{\ast}_{Ne} {\left(A_R^{lN_i}\right)}_{\nu e}
{\delta}_{\Delta\sigma,+1} \sqrt{( 1+2\lambda\beta')
(1-2\bar{\lambda}\beta'') },\\
&& \nonumber \\
A_1^{u_i}\left( \sig;\lam\right)&=&{A_1^{t_i}}^{\ast}\left(\lambda
\longleftrightarrow \bar{\lambda} ,\beta' \longleftrightarrow \beta'' \right),
\end{eqnarray}
\begin{eqnarray}
&&A^{s_i}\left( \sig ; \lam \right) =  \biggl\{ A_L^{l_i}
{\delta}_{\Delta\sigma,-1}+
A_R^{l_i} {\delta}_{\Delta\sigma,+1} \biggr\} \ \ \ \ \ \ \  \\
&&\times \biggl\{ {\left( A_L^{N_i}\right)}_{N\nu}
\sqrt{ ( 1-2\lambda\beta')( 1+2\bar{\lambda}\beta'') }
+  {\left( A_R^{N_i}\right) }_{N\nu}
\sqrt{ ( 1+2\lambda\beta')( 1-2\bar{\lambda}\beta'') } \biggr\} , \nonumber
\end{eqnarray}
where
$$
\left( A_L^{N_i}
\right)_{N\nu}=\frac{g}{2\cos{\Theta_W}}\xi^{(i)}_{LR}\Omega_{N\nu}
\;\;\;\;\;\;\left( A_R^{N_i} \right)_{N\nu}=-\left(
A_L^{N_i}\right)^{\ast}_{N\nu}
\eqno(A.5')
$$
with
$$
\xi^{(1)}_{LR}=\cos{\phi}-\frac{\sin{\phi}}{\sqrt{\cos{2\Theta_W}}},\;\;\;
\xi^{(2)}_{LR}=\sin{\phi}+\frac{\cos{\phi}}{\sqrt{\cos{2\Theta_W}}}\eqno(A.5'')
$$
\begin{eqnarray}
&&A_2^{t_i}=\frac{M_{\nu}M_N}{M_{W_i}^2}  \biggl\{ {\left( A_L^{lN_i}
\right)}_{\nu e} {\delta}_{\sigma ,-1/2}\sqrt{(1+2\lambda\beta' ) }
+ {\left( A_R^{lN_i}
\right)}_{\nu e} {\delta}_{\sigma ,+1/2}\sqrt{(1-
2\lambda\beta' ) } \biggr\}   \nonumber \\
&& \times \biggl\{ {\left( A_L^{lN_i}
\right)}_{Ne}^{\ast} {\delta}_{\bar{\sigma} ,+1/2}\sqrt{(1-
2\bar{\lambda}\beta'' ) }
+ {\left( A_R^{lN_i}
\right)}_{Ne}^{\ast} {\delta}_{\bar{\sigma} ,-1/2}\sqrt{(1+2\bar{\lambda}
\beta'' ) } \biggr\},
\end{eqnarray}
\begin{equation}
A_2^{u_i}\left( \sig;\lam\right)={A_2^{t_i}}\left(\lambda \longleftrightarrow
\bar{\lambda} ,\beta' \longleftrightarrow \beta'',\nu \longleftrightarrow N
\right),
\end{equation}
\begin{eqnarray}
h^{t_i}&=&+2{(-)}^{\sigma -\lambda} \biggl\{
{\left(A_R^{lN_i}\right)}^{\ast}_{Ne} {\left(A_L^{lN_i}\right)}_{\nu e}
{\delta}_{\sigma,-1/2} \sqrt{( 1-2\lambda\beta')
(1-2\lambda\beta'') }  \nonumber \\
&&+ {\left(A_L^{lN_i}\right)}^{\ast}_{Ne} {\left(A_R^{lN_i}\right)}_{\nu e}
{\delta}_{\sigma,+1/2} \sqrt{( 1+2\lambda\beta')
(1+2\lambda\beta'')  } \biggr\}, \nonumber \\
h^{u_i}&=&-h^{t_i}\left( \nu \longleftrightarrow N \right)
\end{eqnarray}

\begin{eqnarray}
&& A^t_{H}=
\biggl( {\left( B_L^{lN} \right)}_{\nu e} {\delta}_{\sigma ,-1/2}\sqrt{1+
2\lambda \beta'  } + {\left( B_R^{lN}
\right)}_{\nu e} {\delta}_{\sigma ,+1/2}\sqrt{(1-2\lambda
\beta' ) } \biggr) \times \nonumber \\
&& \biggl( {\left( B_R^{lN} \right)}_{Ne} {\delta}_{\bar{\sigma} ,
-1/2}\sqrt{1+2\bar{\lambda} \beta''  } + {\left( B_L^{lN}
\right)}_{Ne} {\delta}_{\bar{\sigma} ,+1/2}\sqrt{(1-2\bar{\lambda}
\beta'' ) } \biggr) ,
\end{eqnarray}
\begin{equation}
A^u_{H}\left( \sig;\lam\right)={A^t_{H}}\left(\lambda \longleftrightarrow
\bar{\lambda} ,\beta' \longleftrightarrow \beta'',\nu \longleftrightarrow N
 \right),
\end{equation}

and finally
\begin{equation}
f^{t,u}=\pm \frac12 A_2^{t,u}\left( {(-)}^{\sigma -\lambda} \mp \cos\Theta
\right) ,
\end{equation}
\begin{equation}
g^{t,u}_{H}=\pm \frac12 A^{t,u}_{H}\left( {(-)}^{\sigma -\lambda} \mp
\cos\Theta
\right) ,
\end{equation}
and
\begin{eqnarray}
&&g^s_{H^0}\left( \sig ; \lam \right) = \bar{\sigma} \bar{\lambda}
\biggl\{ \left( B_L^l \right)_{ee}{\delta}_{\sigma,-1/2}-
\left(B_R^l \right)_{ee} {\delta}_{\sigma,+1/2} \biggr\} \ \ \ \ \ \ \ \\
&&                                      \nonumber \\
&&\times \biggl\{ {\left( B_R^N\right)}_{\nu N}
\sqrt{ ( 1-2\lambda\beta')( 1-2\bar{\lambda}\beta'') }
-  {\left( B_L^N\right) }_{\nu N}
\sqrt{ ( 1+2\lambda\beta')( 1+2\bar{\lambda}\beta'') } \biggr\} ,\nonumber
\end{eqnarray}
We present also the vertices that describe the coupling between Higgs and
fermion particles in the left-right symmetric model. In general the vertex
is given by (Fig.3)
\begin{equation}
i\Gamma^{(x)}\left(H\right) \equiv i \left(
B_L^{(x)}\left(H\right)P_L+B_R^{(x)}
\left( H\right)P_R\right),\;\;\;x=l,N,lN.
\end{equation}
Then, in the approximation we make $\left( v_R \gg y \right)$ when we
neglect the electron mass, the left $\left(B_L^{(x)}\right)$ and right $\left(
B_R^{(x)} \right)$ handed couplings for the diagrams from Fig.2 are given by
\begin{itemize}
\item $H_1^{\pm}$ exchange
\end{itemize}
\begin{eqnarray}
\Gamma^{lN}\left( H_1^{\pm} \right)_{N_be^-}&=&\frac{1}{v_R}\sum_{c=4,5,6}
X_{bc}\left( K_L \right)_{ce}P_L, \nonumber \\
\Gamma^{lN}\left( H_1^{\pm} \right)_{e^+N_a}&=&\frac{1}{v_R}\sum_{c=4,5,6}
\left( K_L^{\dagger} \right)_{ec}\left( X^{\ast} \right)_{ca}P_R.
\end{eqnarray}
\begin{itemize}
\item $H_2^{\pm}$ exchange $\left( \alpha_2=\frac{\sqrt{2}}{y\sqrt{1-
\epsilon^2}} \right)$.
\end{itemize}
\begin{eqnarray}
&&\Gamma^{lN}\left( H_2^{\pm} \right)_{N_be^-}=
-\left[ m_b^N\left( K_L\right)_{be}\epsilon\alpha_2 \right]P_L + \\
&&\left[ -\sum_{c=4,5,6} \left(
\Omega_L \right)_{bc}m_c^N \left(K_R\right)_{ce}\alpha_2
+ \frac{1}{\alpha_2v_R^2}\sum_{c=4,5,6} \left(
\Omega_R^{\ast} \right)_{bc}m_c^N \left(K_R\right)_{ce} \right] P_R , \nonumber
\\
&& \nonumber \\
&&\Gamma^{lN}\left( H_2^{\pm} \right)_{e^+N_a}=
-\left[ \left(K_L^{\dagger} \right)_{ea}m_a^N\epsilon\alpha_2 \right]P_R + \\
&& \left[ -\sum_{c=4,5,6} \left(
K_R^{\dagger}\right)_{ec} m_c^N \left( \Omega_L \right)_{ca}\alpha_2
+ \frac{1}{\alpha_2v_R^2}\sum_{c=4,5,6} \left(
K_R^{\dagger} \right)_{ec} m_c^N \left( \Omega_R^{\ast} \right)_{ca} \right]
P_L.
\nonumber
\end{eqnarray}
\begin{itemize}
\item $H_0^0,H_1^0$ and $A_1^0$ exchanges.
\end{itemize}

If we denote
\begin{eqnarray}
A_0&\simeq&\frac{1}{y\sqrt{1-\epsilon^2}}\left[H_1^0-iA_1^0\right], \nonumber
\\
B_0&\simeq& \frac{1}{y}H_0^0-\frac{\epsilon}{y\sqrt{1-\epsilon^2}} \left[
H_1^0+iA_1^0\right],
\end{eqnarray}
then the couplings  $\left(e^-e^+H\right)$ and $\left(N_aN_bH\right)$
can be find from
\begin{eqnarray}
\Gamma^l\left(H_1^0,A_1^0\right)_{e^-e^+}&=&-\left[ \sum_{c=4,5,6}
\left( K_L^{\ast} \right)_{ce} \left( K_R \right)_{ce} m_c^N A_0 \right] P_R
\nonumber \\
&&-\left[ \sum_{c=4,5,6}
\left( K_L\right)_{ce} \left( K_R^{\ast} \right)_{ce} m_c^N A_0^{\ast} \right]
P_L,
\end{eqnarray}
and we can see that the lightest Higgs $H_0^0$ does not couple to $e^-e^+$ (if
we
neglect the electron mass),
\begin{eqnarray}
\Gamma^N\left(H_0^0,H_1^0,A_1^0\right)_{N_aN_b}&=&
-\left[ \left( \left( \Omega_L \right)_{ba}m_a^N+\left(\Omega_L
\right)_{ab} m_b^N \right) B_0 \right.  \\
&+& \left. \sum_{l=\mu,\tau}m_l\left( \left( K_L \right)_{bl}
\left(K_R^{\ast} \right)_{al}+\left( K_L \right)_{al}
\left(K_R^{\ast} \right)_{bl} \right)A_0^{\ast}\right]P_R \nonumber \\
&-&\left[ \left( \left( \Omega_L \right)_{ba}m_a^N+\left(\Omega_L
\right)_{ab} m_b^N \right) B_0^{\ast} \right. \nonumber \\
&+& \left. \sum_{l=\mu,\tau}m_l\left( \left( K_L^{\ast}
\right)_{bl} \left(K_R \right)_{al}+\left( K_L^{\ast} \right)_{al}
\left(K_R \right)_{bl} \right)A_0\right]P_L. \nonumber
\end{eqnarray}
$^{\dag}$ e-mail gluza@usctoux1.cto.us.edu.pl
\newline
$^{\ddagger}$e-mail zralek@usctoux1.cto.us.edu.pl

\section*{Figure Captions}
\newcounter{bean}
\begin{list}
{\bf Fig.\arabic{bean}}{\usecounter{bean}\setlength{\rightmargin}
{\leftmargin}}
\item  Diagrams with gauge boson exchange describing the process $e^-e^+
\rightarrow \nu N$
in the left-right symmetric model on the tree level.
\item  Diagrams with Higgs boson exchange describing the process $e^-e^+
\rightarrow \nu N$
in the left-right symmetric model on the tree level.
\item  Higgs-fermions vertex. l and N denote charged and neutral leptons
respectively.
\item  The $\epsilon$-dependence for $\sigma(e^-e^+ \rightarrow
\nu_{\tau}N(100))$
cross section given by all physical Higgs bosons exchange (solid line) and
only  for  $H_2^+$ Higgs exchange (asterisks line) for LEPII energy.
\item  Total cross section for the single heavy Majorana
and three different light neutrinos production, $\nu_e$ (cross-line),
$\nu_{\mu}$
(dotted line) and $\nu_{\tau}$ (asterisks line) as the energy function. As the
light
neutrinos are not detected only the sum of these three processes can be
observed (solid line).
\end{list}
\newpage
%
%
\input FEYNMAN
\bigphotons
\centerline{Fig.1.}
\begin{picture}(20000,20000)(0,0)
\thicklines
\drawline\fermion[\SE\REG](1500,15000)[5000]
\put(0,\pfronty){$e^-$}
\drawline\photon[\S\REG](\pbackx,\pbacky)[5]
\put(6000,\pmidy){$W_1^+,W_2^+$}
\drawline\fermion[\SW\REG](\photonbackx,\photonbacky)[5000]
\put(0,\pbacky){$e^+$}
\drawline\fermion[\SE\REG](\photonbackx,\photonbacky)[5000]
\put(10000,\pbacky){$N_a$}
\drawline\fermion[\NE\REG](\photonfrontx,\photonfronty)[5000]
\put(10000,\pbacky){$N_b$}
\put(4500,0){(a)}
\end{picture}
\hskip .1cm
\begin{picture}(20000,20000)(0,0)
\thicklines
\drawline\fermion[\SE\REG](1500,15000)[5000]
\put(0,\pfronty){$e^-$}
\global\seglength=1000
\global\gaplength=200
\drawline\photon[\S\REG](\pbackx,\pbacky)[5]
\put(500,\pmidy){$W_1^+,W_2^+$}
\drawline\fermion[\SW\REG](\photonbackx,\photonbacky)[5000]
\put(0,\pbacky){$e^+$}
\drawline\fermion[\NE\REG](\photonbackx,\photonbacky)[12000]
\put(15000,\pbacky){$N_b$}
\drawline\fermion[\SE\REG](\photonfrontx,\photonfronty)[12000]
\put(15000,\pbacky){$N_a$}
\put(7000,0){(b)}
\end{picture}
\begin{picture}(20000,20000)(-8000,-8000)
\thicklines
\drawline\photon[\E\REG](5000,4000)[9]
\put(\pfrontx,5000){$\;\;\;\;\;\;\;Z_1^0,Z_2^0$}
\drawline\fermion[\SW\REG](\photonfrontx,\photonfronty)[5000]
\put(0,\pbacky){$e^+$}
\drawline\fermion[\NW\REG](\photonfrontx,\photonfronty)[5000]
\put(0,\pbacky){$e^-$}
\drawline\fermion[\SE\REG](\photonbackx,\photonbacky)[5000]
\put(19000,\pbacky){$N_a$}
\drawline\fermion[\NE\REG](\photonbackx,\photonbacky)[5000]
\put(19000,\pbacky){$N_b$}
\put(9000,-3000){(c)}
\end{picture}
\newpage
\centerline{Fig.2.}
\begin{picture}(20000,20000)(0,0)
\thicklines
\drawline\fermion[\SE\REG](1500,15000)[5000]
\put(0,\pfronty){$e^-$}
\global\seglength=1000
\global\gaplength=200
\drawline\scalar[\S\REG](\pbackx,\pbacky)[5]
\put(6000,\pmidy){$H_1^+,H_2^+$}
\drawline\fermion[\SW\REG](\scalarbackx,\scalarbacky)[5000]
\put(0,\pbacky){$e^+$}
\drawline\fermion[\SE\REG](\scalarbackx,\scalarbacky)[5000]
\put(10000,\pbacky){$N_a$}
\drawline\fermion[\NE\REG](\scalarfrontx,\scalarfronty)[5000]
\put(10000,\pbacky){$N_b$}
\put(4500,0){(a)}
\end{picture}
\hskip .1cm
\begin{picture}(20000,20000)(0,0)
\thicklines
\drawline\fermion[\SE\REG](1500,15000)[5000]
\put(0,\pfronty){$e^-$}
\global\seglength=1000
\global\gaplength=200
\drawline\scalar[\S\REG](\pbackx,\pbacky)[5]
\put(500,\pmidy){$H_1^+,H_2^+$}
\drawline\fermion[\SW\REG](\scalarbackx,\scalarbacky)[5000]
\put(0,\pbacky){$e^+$}
\drawline\fermion[\NE\REG](\scalarbackx,\scalarbacky)[12000]
\put(15000,\pbacky){$N_b$}
\drawline\fermion[\SE\REG](\scalarfrontx,\scalarfronty)[12000]
\put(15000,\pbacky){$N_a$}
\put(7000,0){(b)}
\end{picture}
\begin{picture}(20000,20000)(-8000,-8000)
\thicklines
\drawline\scalar[\E\REG](5000,5000)[8]
\put(\pfrontx,6000){$\;\;\;\;\;\;\;\; H_1^0,A_1^0$}
\drawline\fermion[\SW\REG](\scalarfrontx,\scalarfronty)[5000]
\put(0,\pbacky){$e^+$}
\drawline\fermion[\NW\REG](\scalarfrontx,\scalarfronty)[5000]
\put(0,\pbacky){$e^-$}
\drawline\fermion[\SE\REG](\scalarbackx,\scalarbacky)[5000]
\put(19000,\pbacky){$N_a$}
\drawline\fermion[\NE\REG](\scalarbackx,\scalarbacky)[5000]
\put(19000,\pbacky){$N_b$}
\put(9000,-3000){(c)}
\end{picture}
\newpage
\centerline{Fig.3.}
\begin{picture}(5000,5000)
\thicklines
\global\seglength=1100
\global\gaplength=350
\drawline\scalar[\E\REG](8000,-3000)[5]
\put(7000,\scalarfronty){H}
\put(16500,\scalarfronty){$i\Gamma^{l,N,lN}\left(H \right)$}
\drawline\fermion[\SE\REG](\scalarbackx,\scalarbacky)[\scalarlengthx]
\put(20500,2500){l,$N_a$,l}
\drawline\fermion[\NE\REG](\scalarbackx,\scalarbacky)[\scalarlengthx]
\put(20500,-9000){l',$N_b$,N}
\advance\pbackx by -5000
\advance\pbacky by -5000
\end{picture}
\newpage

%
%
\pagenumbering{25}
\setlength{\unitlength}{0.240900pt}
\ifx\plotpoint\undefined\newsavebox{\plotpoint}\fi
\sbox{\plotpoint}{\rule[-0.200pt]{0.400pt}{0.400pt}}%
\begin{picture}(1500,900)(0,0)
\font\gnuplot=cmr10 at 10pt
\gnuplot
\sbox{\plotpoint}{\rule[-0.200pt]{0.400pt}{0.400pt}}%
\put(220.0,113.0){\rule[-0.200pt]{0.400pt}{184.048pt}}
\put(220.0,113.0){\rule[-0.200pt]{4.818pt}{0.400pt}}
\put(198,113){\makebox(0,0)[r]{1e-10}}
\put(1416.0,113.0){\rule[-0.200pt]{4.818pt}{0.400pt}}
\put(220.0,151.0){\rule[-0.200pt]{2.409pt}{0.400pt}}
\put(1426.0,151.0){\rule[-0.200pt]{2.409pt}{0.400pt}}
\put(220.0,202.0){\rule[-0.200pt]{2.409pt}{0.400pt}}
\put(1426.0,202.0){\rule[-0.200pt]{2.409pt}{0.400pt}}
\put(220.0,228.0){\rule[-0.200pt]{2.409pt}{0.400pt}}
\put(1426.0,228.0){\rule[-0.200pt]{2.409pt}{0.400pt}}
\put(220.0,240.0){\rule[-0.200pt]{4.818pt}{0.400pt}}
\put(198,240){\makebox(0,0)[r]{1e-09}}
\put(1416.0,240.0){\rule[-0.200pt]{4.818pt}{0.400pt}}
\put(220.0,279.0){\rule[-0.200pt]{2.409pt}{0.400pt}}
\put(1426.0,279.0){\rule[-0.200pt]{2.409pt}{0.400pt}}
\put(220.0,329.0){\rule[-0.200pt]{2.409pt}{0.400pt}}
\put(1426.0,329.0){\rule[-0.200pt]{2.409pt}{0.400pt}}
\put(220.0,355.0){\rule[-0.200pt]{2.409pt}{0.400pt}}
\put(1426.0,355.0){\rule[-0.200pt]{2.409pt}{0.400pt}}
\put(220.0,368.0){\rule[-0.200pt]{4.818pt}{0.400pt}}
\put(198,368){\makebox(0,0)[r]{1e-08}}
\put(1416.0,368.0){\rule[-0.200pt]{4.818pt}{0.400pt}}
\put(220.0,406.0){\rule[-0.200pt]{2.409pt}{0.400pt}}
\put(1426.0,406.0){\rule[-0.200pt]{2.409pt}{0.400pt}}
\put(220.0,457.0){\rule[-0.200pt]{2.409pt}{0.400pt}}
\put(1426.0,457.0){\rule[-0.200pt]{2.409pt}{0.400pt}}
\put(220.0,483.0){\rule[-0.200pt]{2.409pt}{0.400pt}}
\put(1426.0,483.0){\rule[-0.200pt]{2.409pt}{0.400pt}}
\put(220.0,495.0){\rule[-0.200pt]{4.818pt}{0.400pt}}
\put(198,495){\makebox(0,0)[r]{1e-07}}
\put(1416.0,495.0){\rule[-0.200pt]{4.818pt}{0.400pt}}
\put(220.0,533.0){\rule[-0.200pt]{2.409pt}{0.400pt}}
\put(1426.0,533.0){\rule[-0.200pt]{2.409pt}{0.400pt}}
\put(220.0,584.0){\rule[-0.200pt]{2.409pt}{0.400pt}}
\put(1426.0,584.0){\rule[-0.200pt]{2.409pt}{0.400pt}}
\put(220.0,610.0){\rule[-0.200pt]{2.409pt}{0.400pt}}
\put(1426.0,610.0){\rule[-0.200pt]{2.409pt}{0.400pt}}
\put(220.0,622.0){\rule[-0.200pt]{4.818pt}{0.400pt}}
\put(198,622){\makebox(0,0)[r]{1e-06}}
\put(1416.0,622.0){\rule[-0.200pt]{4.818pt}{0.400pt}}
\put(220.0,661.0){\rule[-0.200pt]{2.409pt}{0.400pt}}
\put(1426.0,661.0){\rule[-0.200pt]{2.409pt}{0.400pt}}
\put(220.0,711.0){\rule[-0.200pt]{2.409pt}{0.400pt}}
\put(1426.0,711.0){\rule[-0.200pt]{2.409pt}{0.400pt}}
\put(220.0,737.0){\rule[-0.200pt]{2.409pt}{0.400pt}}
\put(1426.0,737.0){\rule[-0.200pt]{2.409pt}{0.400pt}}
\put(220.0,750.0){\rule[-0.200pt]{4.818pt}{0.400pt}}
\put(198,750){\makebox(0,0)[r]{1e-05}}
\put(1416.0,750.0){\rule[-0.200pt]{4.818pt}{0.400pt}}
\put(220.0,788.0){\rule[-0.200pt]{2.409pt}{0.400pt}}
\put(1426.0,788.0){\rule[-0.200pt]{2.409pt}{0.400pt}}
\put(220.0,839.0){\rule[-0.200pt]{2.409pt}{0.400pt}}
\put(1426.0,839.0){\rule[-0.200pt]{2.409pt}{0.400pt}}
\put(220.0,865.0){\rule[-0.200pt]{2.409pt}{0.400pt}}
\put(1426.0,865.0){\rule[-0.200pt]{2.409pt}{0.400pt}}
\put(220.0,877.0){\rule[-0.200pt]{4.818pt}{0.400pt}}
\put(198,877){\makebox(0,0)[r]{0.0001}}
\put(1416.0,877.0){\rule[-0.200pt]{4.818pt}{0.400pt}}
\put(220.0,113.0){\rule[-0.200pt]{0.400pt}{4.818pt}}
\put(220,68){\makebox(0,0){0}}
\put(220.0,857.0){\rule[-0.200pt]{0.400pt}{4.818pt}}
\put(463.0,113.0){\rule[-0.200pt]{0.400pt}{4.818pt}}
\put(463,68){\makebox(0,0){0.2}}
\put(463.0,857.0){\rule[-0.200pt]{0.400pt}{4.818pt}}
\put(706.0,113.0){\rule[-0.200pt]{0.400pt}{4.818pt}}
\put(706,68){\makebox(0,0){0.4}}
\put(706.0,857.0){\rule[-0.200pt]{0.400pt}{4.818pt}}
\put(950.0,113.0){\rule[-0.200pt]{0.400pt}{4.818pt}}
\put(950,68){\makebox(0,0){0.6}}
\put(950.0,857.0){\rule[-0.200pt]{0.400pt}{4.818pt}}
\put(1193.0,113.0){\rule[-0.200pt]{0.400pt}{4.818pt}}
\put(1193,68){\makebox(0,0){0.8}}
\put(1193.0,857.0){\rule[-0.200pt]{0.400pt}{4.818pt}}
\put(1436.0,113.0){\rule[-0.200pt]{0.400pt}{4.818pt}}
\put(1436,68){\makebox(0,0){1}}
\put(1436.0,857.0){\rule[-0.200pt]{0.400pt}{4.818pt}}
\put(220.0,113.0){\rule[-0.200pt]{292.934pt}{0.400pt}}
\put(1436.0,113.0){\rule[-0.200pt]{0.400pt}{184.048pt}}
\put(220.0,877.0){\rule[-0.200pt]{292.934pt}{0.400pt}}
\put(45,495){\makebox(0,0){$\sigma$ [fb]}}
\put(828,23){\makebox(0,0){$\epsilon$}}
\put(828,-5000){\makebox(0,0){numb}}
\put(828,950){\makebox(0,0){Fig.4}}
\put(220.0,113.0){\rule[-0.200pt]{0.400pt}{184.048pt}}
\put(1306,700){\makebox(0,0)[r]{$\sigma$(all Higgses)}}
\put(1328.0,700.0){\rule[-0.200pt]{15.899pt}{0.400pt}}
\put(220,824){\usebox{\plotpoint}}
\put(220.0,824.0){\rule[-0.200pt]{292.934pt}{0.400pt}}
\put(1306,640){\makebox(0,0)[r]{$\sigma$($H_2^{\pm}$)}}
\put(1350,640){\raisebox{-.8pt}{\makebox(0,0){$\ast$}}}
\put(220,143){\raisebox{-.8pt}{\makebox(0,0){$\ast$}}}
\put(281,145){\raisebox{-.8pt}{\makebox(0,0){$\ast$}}}
\put(342,152){\raisebox{-.8pt}{\makebox(0,0){$\ast$}}}
\put(402,162){\raisebox{-.8pt}{\makebox(0,0){$\ast$}}}
\put(463,174){\raisebox{-.8pt}{\makebox(0,0){$\ast$}}}
\put(524,187){\raisebox{-.8pt}{\makebox(0,0){$\ast$}}}
\put(585,200){\raisebox{-.8pt}{\makebox(0,0){$\ast$}}}
\put(646,214){\raisebox{-.8pt}{\makebox(0,0){$\ast$}}}
\put(706,227){\raisebox{-.8pt}{\makebox(0,0){$\ast$}}}
\put(767,240){\raisebox{-.8pt}{\makebox(0,0){$\ast$}}}
\put(828,252){\raisebox{-.8pt}{\makebox(0,0){$\ast$}}}
\put(889,265){\raisebox{-.8pt}{\makebox(0,0){$\ast$}}}
\put(950,279){\raisebox{-.8pt}{\makebox(0,0){$\ast$}}}
\put(1010,292){\raisebox{-.8pt}{\makebox(0,0){$\ast$}}}
\put(1071,307){\raisebox{-.8pt}{\makebox(0,0){$\ast$}}}
\put(1132,322){\raisebox{-.8pt}{\makebox(0,0){$\ast$}}}
\put(1193,340){\raisebox{-.8pt}{\makebox(0,0){$\ast$}}}
\put(1254,361){\raisebox{-.8pt}{\makebox(0,0){$\ast$}}}
\put(1314,388){\raisebox{-.8pt}{\makebox(0,0){$\ast$}}}
\put(1375,431){\raisebox{-.8pt}{\makebox(0,0){$\ast$}}}
\put(1436,524){\raisebox{-.8pt}{\makebox(0,0){$\ast$}}}
\end{picture}
\vskip .5in

\setlength{\unitlength}{0.240900pt}
\ifx\plotpoint\undefined\newsavebox{\plotpoint}\fi
\sbox{\plotpoint}{\rule[-0.200pt]{0.400pt}{0.400pt}}%
\begin{picture}(1500,900)(0,0)
\font\gnuplot=cmr10 at 10pt
\gnuplot
\sbox{\plotpoint}{\rule[-0.200pt]{0.400pt}{0.400pt}}%
\put(220.0,113.0){\rule[-0.200pt]{292.934pt}{0.400pt}}
\put(220.0,113.0){\rule[-0.200pt]{4.818pt}{0.400pt}}
\put(198,113){\makebox(0,0)[r]{0}}
\put(1416.0,113.0){\rule[-0.200pt]{4.818pt}{0.400pt}}
\put(220.0,189.0){\rule[-0.200pt]{4.818pt}{0.400pt}}
\put(198,189){\makebox(0,0)[r]{1}}
\put(1416.0,189.0){\rule[-0.200pt]{4.818pt}{0.400pt}}
\put(220.0,266.0){\rule[-0.200pt]{4.818pt}{0.400pt}}
\put(198,266){\makebox(0,0)[r]{2}}
\put(1416.0,266.0){\rule[-0.200pt]{4.818pt}{0.400pt}}
\put(220.0,342.0){\rule[-0.200pt]{4.818pt}{0.400pt}}
\put(198,342){\makebox(0,0)[r]{3}}
\put(1416.0,342.0){\rule[-0.200pt]{4.818pt}{0.400pt}}
\put(220.0,419.0){\rule[-0.200pt]{4.818pt}{0.400pt}}
\put(198,419){\makebox(0,0)[r]{4}}
\put(1416.0,419.0){\rule[-0.200pt]{4.818pt}{0.400pt}}
\put(220.0,495.0){\rule[-0.200pt]{4.818pt}{0.400pt}}
\put(198,495){\makebox(0,0)[r]{5}}
\put(1416.0,495.0){\rule[-0.200pt]{4.818pt}{0.400pt}}
\put(220.0,571.0){\rule[-0.200pt]{4.818pt}{0.400pt}}
\put(198,571){\makebox(0,0)[r]{6}}
\put(1416.0,571.0){\rule[-0.200pt]{4.818pt}{0.400pt}}
\put(220.0,648.0){\rule[-0.200pt]{4.818pt}{0.400pt}}
\put(198,648){\makebox(0,0)[r]{7}}
\put(1416.0,648.0){\rule[-0.200pt]{4.818pt}{0.400pt}}
\put(220.0,724.0){\rule[-0.200pt]{4.818pt}{0.400pt}}
\put(198,724){\makebox(0,0)[r]{8}}
\put(1416.0,724.0){\rule[-0.200pt]{4.818pt}{0.400pt}}
\put(220.0,801.0){\rule[-0.200pt]{4.818pt}{0.400pt}}
\put(198,801){\makebox(0,0)[r]{9}}
\put(1416.0,801.0){\rule[-0.200pt]{4.818pt}{0.400pt}}
\put(220.0,877.0){\rule[-0.200pt]{4.818pt}{0.400pt}}
\put(198,877){\makebox(0,0)[r]{10}}
\put(1416.0,877.0){\rule[-0.200pt]{4.818pt}{0.400pt}}
\put(220.0,113.0){\rule[-0.200pt]{0.400pt}{4.818pt}}
\put(220,68){\makebox(0,0){100}}
\put(220.0,857.0){\rule[-0.200pt]{0.400pt}{4.818pt}}
\put(355.0,113.0){\rule[-0.200pt]{0.400pt}{4.818pt}}
\put(355,68){\makebox(0,0){150}}
\put(355.0,857.0){\rule[-0.200pt]{0.400pt}{4.818pt}}
\put(490.0,113.0){\rule[-0.200pt]{0.400pt}{4.818pt}}
\put(490,68){\makebox(0,0){200}}
\put(490.0,857.0){\rule[-0.200pt]{0.400pt}{4.818pt}}
\put(625.0,113.0){\rule[-0.200pt]{0.400pt}{4.818pt}}
\put(625,68){\makebox(0,0){250}}
\put(625.0,857.0){\rule[-0.200pt]{0.400pt}{4.818pt}}
\put(760.0,113.0){\rule[-0.200pt]{0.400pt}{4.818pt}}
\put(760,68){\makebox(0,0){300}}
\put(760.0,857.0){\rule[-0.200pt]{0.400pt}{4.818pt}}
\put(896.0,113.0){\rule[-0.200pt]{0.400pt}{4.818pt}}
\put(896,68){\makebox(0,0){350}}
\put(896.0,857.0){\rule[-0.200pt]{0.400pt}{4.818pt}}
\put(1031.0,113.0){\rule[-0.200pt]{0.400pt}{4.818pt}}
\put(1031,68){\makebox(0,0){400}}
\put(1031.0,857.0){\rule[-0.200pt]{0.400pt}{4.818pt}}
\put(1166.0,113.0){\rule[-0.200pt]{0.400pt}{4.818pt}}
\put(1166,68){\makebox(0,0){450}}
\put(1166.0,857.0){\rule[-0.200pt]{0.400pt}{4.818pt}}
\put(1301.0,113.0){\rule[-0.200pt]{0.400pt}{4.818pt}}
\put(1301,68){\makebox(0,0){500}}
\put(1301.0,857.0){\rule[-0.200pt]{0.400pt}{4.818pt}}
\put(1436.0,113.0){\rule[-0.200pt]{0.400pt}{4.818pt}}
\put(1436,68){\makebox(0,0){550}}
\put(1436.0,857.0){\rule[-0.200pt]{0.400pt}{4.818pt}}
\put(220.0,113.0){\rule[-0.200pt]{292.934pt}{0.400pt}}
\put(1436.0,113.0){\rule[-0.200pt]{0.400pt}{184.048pt}}
\put(220.0,877.0){\rule[-0.200pt]{292.934pt}{0.400pt}}
\put(45,495){\makebox(0,0){$\sigma$ [fb]}}
\put(828,23){\makebox(0,0){$\sqrt{s}$ [GeV]}}
\put(828,950){\makebox(0,0){Fig.5}}
\put(220.0,113.0){\rule[-0.200pt]{0.400pt}{184.048pt}}
\put(1306,710){\makebox(0,0)[r]{$\rightarrow (\nu_e+\nu_{\mu}+
\nu_{\tau})N(100)$}}
\put(1328.0,710){\rule[-0.200pt]{15.899pt}{0.400pt}}
\put(220,113){\usebox{\plotpoint}}
\multiput(220.58,113.00)(0.499,0.723){213}{\rule{0.120pt}{0.678pt}}
\multiput(219.17,113.00)(108.000,154.593){2}{\rule{0.400pt}{0.339pt}}
\multiput(328.58,269.00)(0.499,0.597){213}{\rule{0.120pt}{0.578pt}}
\multiput(327.17,269.00)(108.000,127.801){2}{\rule{0.400pt}{0.289pt}}
\multiput(436.58,398.00)(0.499,0.667){213}{\rule{0.120pt}{0.633pt}}
\multiput(435.17,398.00)(108.000,142.685){2}{\rule{0.400pt}{0.317pt}}
\multiput(544.00,542.58)(0.643,0.499){165}{\rule{0.614pt}{0.120pt}}
\multiput(544.00,541.17)(106.725,84.000){2}{\rule{0.307pt}{0.400pt}}
\multiput(652.00,626.58)(0.872,0.499){121}{\rule{0.797pt}{0.120pt}}
\multiput(652.00,625.17)(106.346,62.000){2}{\rule{0.398pt}{0.400pt}}
\multiput(760.00,688.58)(1.189,0.498){89}{\rule{1.048pt}{0.120pt}}
\multiput(760.00,687.17)(106.825,46.000){2}{\rule{0.524pt}{0.400pt}}
\multiput(869.00,734.58)(1.468,0.498){71}{\rule{1.268pt}{0.120pt}}
\multiput(869.00,733.17)(105.369,37.000){2}{\rule{0.634pt}{0.400pt}}
\multiput(977.00,771.58)(1.814,0.497){57}{\rule{1.540pt}{0.120pt}}
\multiput(977.00,770.17)(104.804,30.000){2}{\rule{0.770pt}{0.400pt}}
\multiput(1085.00,801.58)(2.019,0.497){51}{\rule{1.700pt}{0.120pt}}
\multiput(1085.00,800.17)(104.472,27.000){2}{\rule{0.850pt}{0.400pt}}
\multiput(1193.00,828.58)(3.051,0.495){33}{\rule{2.500pt}{0.119pt}}
\multiput(1193.00,827.17)(102.811,18.000){2}{\rule{1.250pt}{0.400pt}}
\put(1306,660){\makebox(0,0)[r]{$\rightarrow \nu_{\tau}N(100)$}}
\put(1350,660){\raisebox{-.8pt}{\makebox(0,0){$\ast$}}}
\put(220,113){\raisebox{-.8pt}{\makebox(0,0){$\ast$}}}
\put(328,164){\raisebox{-.8pt}{\makebox(0,0){$\ast$}}}
\put(436,217){\raisebox{-.8pt}{\makebox(0,0){$\ast$}}}
\put(545,255){\raisebox{-.8pt}{\makebox(0,0){$\ast$}}}
\put(653,283){\raisebox{-.8pt}{\makebox(0,0){$\ast$}}}
\put(761,303){\raisebox{-.8pt}{\makebox(0,0){$\ast$}}}
\put(869,319){\raisebox{-.8pt}{\makebox(0,0){$\ast$}}}
\put(977,331){\raisebox{-.8pt}{\makebox(0,0){$\ast$}}}
\put(1085,341){\raisebox{-.8pt}{\makebox(0,0){$\ast$}}}
\put(1193,349){\raisebox{-.8pt}{\makebox(0,0){$\ast$}}}
\put(1301,356){\raisebox{-.8pt}{\makebox(0,0){$\ast$}}}
\sbox{\plotpoint}{\rule[-0.400pt]{0.800pt}{0.800pt}}%
\put(1306,620){\makebox(0,0)[r]{$\rightarrow \nu_{\mu}N(100)$}}
\put(1350,620){\rule{1pt}{1pt}}
\put(220,113){\rule{1pt}{1pt}}
\put(328,136){\rule{1pt}{1pt}}
\put(436,161){\rule{1pt}{1pt}}
\put(545,179){\rule{1pt}{1pt}}
\put(653,192){\rule{1pt}{1pt}}
\put(761,202){\rule{1pt}{1pt}}
\put(869,210){\rule{1pt}{1pt}}
\put(977,216){\rule{1pt}{1pt}}
\put(1085,220){\rule{1pt}{1pt}}
\put(1193,224){\rule{1pt}{1pt}}
\put(1301,227){\rule{1pt}{1pt}}
\sbox{\plotpoint}{\rule[-0.500pt]{1.000pt}{1.000pt}}%
\put(1306,570){\makebox(0,0)[r]{$\rightarrow \nu_e N(100)$}}
\multiput(1328,570)(20.756,0.000){4}{\usebox{\plotpoint}}
\put(1394,570){\usebox{\plotpoint}}
\put(220,113){\usebox{\plotpoint}}
\multiput(220,113)(16.604,12.453){7}{\usebox{\plotpoint}}
\multiput(328,194)(16.531,12.551){7}{\usebox{\plotpoint}}
\multiput(436,276)(18.323,9.750){5}{\usebox{\plotpoint}}
\multiput(545,334)(19.283,7.678){6}{\usebox{\plotpoint}}
\multiput(653,377)(19.950,5.726){6}{\usebox{\plotpoint}}
\multiput(761,408)(20.261,4.503){5}{\usebox{\plotpoint}}
\multiput(869,432)(20.473,3.412){5}{\usebox{\plotpoint}}
\multiput(977,450)(20.558,2.855){5}{\usebox{\plotpoint}}
\multiput(1085,465)(20.629,2.292){6}{\usebox{\plotpoint}}
\multiput(1193,477)(20.684,1.724){5}{\usebox{\plotpoint}}
\put(1301,486){\usebox{\plotpoint}}
\put(1350,570){\makebox(0,0){$+$}}
\put(220,113){\makebox(0,0){$+$}}
\put(328,194){\makebox(0,0){$+$}}
\put(436,276){\makebox(0,0){$+$}}
\put(545,334){\makebox(0,0){$+$}}
\put(653,377){\makebox(0,0){$+$}}
\put(761,408){\makebox(0,0){$+$}}
\put(869,432){\makebox(0,0){$+$}}
\put(977,450){\makebox(0,0){$+$}}
\put(1085,465){\makebox(0,0){$+$}}
\put(1193,477){\makebox(0,0){$+$}}
\put(1301,486){\makebox(0,0){$+$}}
\end{picture}


\begin{thebibliography}{19}

\bibitem{1intr} J.C.Pati and A.Salam, Phys.Rev.D{\bf 10},275(1974);
R.N.Mohapatra and J.C.Pati,ibid.{\bf 11},566(1975);{\bf 11},2559(1975);
\newline
G.Senjanovic and R.N.Mohapatra,ibid.{\bf 12},152(1975);G.Senjanovic,
Nucl.Phys.{\bf B153},334(1979).
\bibitem{2intr} N.G.Deshpande,J.F.Gunion,B.Kayser,F.Olness \newline
Phys.Rev.D{\bf 44},837(1991).
\bibitem{3intr} T.Yanagida, Proc.Workshop on Unified Theory and Baryon
number of Universe, ed. by Sawada and Sugamoto (KEK,1979),
M.Gell-Mann,P.Ramond and R.Slansky, in Supergravity, ed.by van
Nieuwenhuizen and Freedman (North-Holland,Amsterdam,1979).
\bibitem{4intr} J.Gluza and M.Zra{\l}ek,Phys.Rev.D{\bf 48},5093(1993).
\bibitem{5intr} R.Vuopionpera,"Production of Heavy Neutrinos in $e^+e^-$
Collisions",Report series HU-SEFT R 1993-04.
\bibitem{zralek} J.Polak and M.Zra{\l}ek,Nucl.Phys.{\bf B363},385(1991);
\newline
Phys.Lett.{\bf B276},492(1992);Phys.Rev.{\bf D46},3871(1992).
\bibitem{sant} A.Santamaria Phys.Lett {\bf B}305,90(1993).
\bibitem{buchpil} W.Buchmuller,C.Greub,Phys.Letters{\bf B256},465(1991);
Nucl.Phys.{\bf B363},345(1991);A.Pilaftsis,Z.Phys.C55,275(1992).
\item G.Beall,M.Bander and A.Soni Phys.Rev.Lett.{\bf 48},848(1982).
\item G.Ecker,W.Grimus and H.Newfeld Phys.Lett.{\bf B127},305(1983);\newline
M.R.Mohapatra,G.Senjanovic and M.D.Tran Phys.Rev.D{\bf 28},546(1983);
\newline F.J.Gilman
and M.H.Reno,ibid,{\bf 29}937(1984);F.I.Olness and M.E.Ebel Phys.Rev.D{\bf 30}
,1034(1984).

\end{thebibliography}
\end{document}